\RequirePackage{ifpdf}
\ifpdf 
\documentclass[pdftex]{sigma}
\else
\documentclass{sigma}
\fi

\begin{document}

\renewcommand{\PaperNumber}{028}

\FirstPageHeading

\renewcommand{\thefootnote}{$\star$}

\ShortArticleName{Noncommutative Lagrange Mechanics}

\ArticleName{Noncommutative Lagrange Mechanics\footnote{This paper is a
contribution to the Proceedings of the 3-rd Microconference
``Analytic and Algebraic Me\-thods~III''. The full collection is
available at
\href{http://www.emis.de/journals/SIGMA/Prague2007.html}{http://www.emis.de/journals/SIGMA/Prague2007.html}}}

\Author{Denis KOCHAN~$^{\dag\ddag}$}

\AuthorNameForHeading{D. Kochan}

\Address{$^\dag$~Dept.~of Theoretical Physics, FMFI UK, Mlynsk\' a dolina F2, 842 48 Bratislava, Slovakia}

\Address{$^\ddag$~Dept.~of Theoretical Physics, Nuclear Physics Institute AS CR, 250 68 \v{R}e\v{z}, Czech Republic}

\EmailD{\href{mailto:denis.kochan@gmail.com}{denis.kochan@gmail.com}}

\ArticleDates{Received November 26, 2007, in f\/inal form January
29, 2008; Published online February 25, 2008}

\Abstract{It is proposed how to impose a general type of ``noncommutativity'' within classical mechanics
from f\/irst principles. Formulation is performed in completely alternative way, i.e.\ without any resort to fuzzy and/or
star product philosophy, which are extensively applied within noncommutative quantum theories.
Newton--Lagrange noncommutative equations of motion are formulated and their properties are analyzed from the pure
geometrical point of view. It is argued that the dynamical quintessence of the system consists in its kinetic energy
(Riemannian metric) specifying Riemann--Levi-Civita connection and thus the \emph{inertia} geodesics of the free motion.
Throughout the paper, ``noncommutativity'' is considered as an internal geometric structure of the
conf\/iguration space, which can not be ``observed'' \emph{per~se}. Manifestation of the noncommutative phenomena is
mediated by the interaction of the system with noncommutative background under the consideration. The simplest model of
the interaction (minimal coupling) is proposed and it is shown that guiding af\/f\/ine connection is modif\/ied by the quadratic
analog of the Lorentz electromagnetic force (contortion term).}

\Keywords{noncommutative mechanics; af\/f\/ine connection; contortion}

\Classification{70G45; 46L55; 53B05}

\rightline{\it Dedicated to my daughter Zoe on the occasion of her birth}

\renewcommand{\thefootnote}{\arabic{footnote}}
\setcounter{footnote}{0}

\section{Introduction and motivation}

Physical models involving ``noncommutativity'' have become very popular and have been extensively studied in
the last decade.
Strong impact comes from string theory. It is well known how the noncommutativity appears in D-branes, when
the open string dynamics is analyzed in the presence of constant B-f\/ield \cite{seiberg}.
Another plumbless source of the ``noncommutativity'' is pure mathematics. New mathematical ideas and structures
are applicable when speculating about the microscopical structure of the space-time.
There is a common belief that this structure is smashed by some fundamental uncertainty and that precise localization
of its events is unreachab\-le~\cite{doplicher}. The concept of the point is lacking sense
and physics is modeled on ``noncommutative space-time.''
The mathematical language of noncommutative physics shifts from visual geometry to rather abstract algebra. The
focus of \emph{Noncommutative Geometry} is nowadays really wide. It includes quantum groups, K-theory and Fredholm
Modules, (cyclic) cohomology and index theory, deformation quantization, fuzzy geometry and so on \cite{madore}.

There is a lot of articles and dissertations elaborating the ongoing noncommutative business. Since my aim is not to
trace out its history and all detailed circumstances, I refer here to survey articles \cite{douglas-szabo}
that will certainly f\/ill this gap (see also patulous reference lists therein).

Noncommutative quantum mechanics is the subject of studies at \cite{chujkovia1,chujkovia2}. The common point of
the f\/irst group of papers lies in simple replacement of the canonical Poisson brackets on the phase space (cotangent
bundle) by modif\/ied NC-brackets. Switching from Classical to Quantum Mechanics is performed in the spirit of Dirac's
canonical quantization dictum or via the path integral approach. However, in~\cite{chujkovia2}, there are proposed
various f\/ield theoretical models which yield, under some circumstances, to the ef\/fective quantum dynamics
corresponding to the motion on the noncommutative (Moyal) plane.

This article is about a similar subject, but noncommutativity is handled from the classical perspective only.
Some ef\/fort concerning this direction was done in \cite{chujkovia3}. Authors of those papers consider noncommutative
classical mechanics in the fashion of symplectic and/or Poisson geometry. They substitute the canonical two-form (Poisson
brackets) on the phase space by modif\/ied quantity, which appears in the dequantization limit $\hbar\rightarrow 0$ in the
Weyl--Wigner--Moyal star product. What they call as noncommutative mechanics is the standard classical mechanics in the
Hamilton's picture, but with the ``modif\/ied'' symplectic two-form. The main objection to such ``noncommutative
treatment'' is that according to Darboux theorem all symplectic structures are locally equivalent, i.e.\ when
performing a suitable change of coordinates one gets from the canonical two-form endowing the phase space the two-form,
which comes from the Weyl--Wigner--Moyal ``dequantization''. Therefore noncommutative dynamical equations derived in~\cite{chujkovia3} are ordinary Hamilton equation of motion, but they are only rewritten in dif\/ferent coordinates. The only
essence would be, if one would be able to canonize some class of coordinates as special and physically privileged. But than
the covariance and treasury of the coordinate free formulation of Hamilton mechanics become broken and results start to
depend on the observer.

The following paper, therefore, deals noncommutative ef\/fects in completely alternative way, i.e.\ without any resorts to
its Hamiltonian and/or Lagrangian precursors. Our starting point is dynamics which is represented by classical equations
of motion and its relation to an af\/f\/ine connection on a underlying conf\/iguration space $M$.
This connection can be specif\/ied by kinetic energy which def\/ines the system and by external forces which produce a
disturbance from the free geodesic motion. The noncommutativity is imposed as an additional internal Poisson structure (non-constant bi-vector) on
the conf\/iguration manifold $M$. Its presence together with the kinetic energy gives rise to a natural nonzero contortion
term. From the point of view of ``commutative RLC-connection'' this term can be interpreted as a background
noncommutative ``Lorentz-like'' force which af\/fects the parallel transport and thus the free motion of the system.
It will be shown that this new contortion force is always perpendicular to the actual velocity and therefore the total
mechanical energy is conserved. Moreover, one can convince him/her-self that this extra noncommutative background
force f\/ield can not be derived from a potential energy function. Therefore the concept of Lagrangian/Hamiltonian is completely missing and there
rises a natural question what will be a reasonable quantum analog of the proposed classical mechanics whose dynamics is
governed by the Newton--Lagrange noncommutative equations of motion. It will be clear that the Moyal--Kontsevich star
product which
can be considered due to present (non-constant) Poisson structure on $M$ does not provide satisfying answer.

\section{Preliminaries}

In the following section we recall some elementary geometry
picturing Lagrange mechanics \cite{arnold}.

An autonomous mechanical system with $m$ degrees of freedom occupies conf\/iguration
space, a $m$-dimensional smooth manifold $M$. Its dynamics is
governed by the kinetic energy (non-singular, positive) tensor $g\in\Gamma(S^2\,T^*M)$
and the co-vectorial strength f\/ield $Q\in\Gamma(T^*M)$ (in general velocity dependent) describing
forces that act within the system:
\[
\frac{d}{dt}\left(\frac{\partial g}{\partial \dot{x}^a}\right)-\frac{\partial g}{\partial x^a}=Q_a, \quad
a=1,\dots, m,\qquad \mbox{and}\qquad g(x,\dot{x})=\tfrac{1}{2} g_{kl}(x)\dot{x}^k\dot{x}^l .
\]
In the special case of exact strength $Q=-dU$ (with potential $U\in C^\infty(M)$),
one can introduce the Lagrangian function $\mathrm{L}: TM\rightarrow \mathbb{R}$:
\[
\bigl(p\in M, v\in T_p M \bigr)\in TM\longmapsto \mathrm{L}(p,v):=g_p(v,v)-U(p)\in\mathbb{R}
\]
and write down the celebrated Euler--Lagrange equations.

Natural setting for classical mechanics is a tangent bundle of the conf\/iguration
space $M$. Dynamics with a given initial position and velocity is then determined by
a symmetric af\/f\/ine connection on $M$ associated with the Riemannian kinetic tensor f\/ield~$g$.

Let me remind the reader how it works. A general af\/f\/ine connection on $M$ is def\/ined as
a~$m$-dimensional horizontal distribution on $TM\xrightarrow{\tau}M$, $(p,v)\overset{\tau}{\mapsto} p$,
which is invariant under the action of scalar multiplication in f\/ibers: $(p,v)\overset{\alpha}{\mapsto}(p,\alpha\cdot v)$
$(\alpha\in\mathbb{R})$. Changed into the coins:
\[
\forall\; (p,v)\in TM\qquad T_{(p,v)}\bigl(TM\bigr)=\mathrm{Ver}_{(p,v)}\oplus\mathrm{Hor}_{(p,v)},
\qquad \mbox{and}\qquad \mathrm{Ver}_{(p,v)} \cap \mathrm{Hor}_{(p,v)}=\bigl\{0\bigr\}.
\]
Using a local coordinate chart $(x^a,\dot{x}^a)$ on a patch of $TM$, one can span the space $\mathrm{Hor}_{(p,v)}$
by a~collection of $m$ vectors:
\[
\partial_{x^a}\biggr|_{(p,v)}^{\mathrm{Hor}}:=\frac{\partial}{\partial x^a}\biggr|_{(p,v)}- A_{a}^b(x,\dot{x}) \frac{\partial}{\partial \dot{x}^b}\biggr|_{(p,v)}.
\]
Requiring the invariance w.r.t.\ the mentioned $\alpha$-scaling:
$\alpha_*\bigl(\mathrm{Hor}_{(p,v)}\bigr)=\mathrm{Hor}_{(p, \alpha\cdot v)}$, we conclude that
$A_a^b(x,\dot{x})=\dot{x}^c\varGamma_{ca}^b(x)$. Here $\varGamma$ stands for a set
of $m^3$ Chris\-tof\-fel symbols.

Prescribing an af\/f\/ine connection on $M$, one is able to perform a horizontal lift of smooth curve
$\gamma: [0,1]\rightarrow M$ to a tangent bundle curve $\widehat{\gamma}: [0,1]\rightarrow TM$.
Concisely, suppose that at a~given time $t$ we are occupying the tangent bundle point $\widehat{\gamma}(t)=(p,v)$ and
a tangent vector to~$\gamma(t)$ at $p=\tau(p,v)$ is $\dot{\gamma}(p)\in T_pM$. Then, after lapsing
an inf\/initesimally short time $\varepsilon$, the new tangent bundle position $\widehat{\gamma}(t+\varepsilon)$
will be specif\/ied by an $\varepsilon$-step in the direction of the horizontal lift $\dot{\gamma}(p)\bigr|_{(p,v)}^{\mathrm{Hor}}$.
In other words, in coordinate patch $(x^a,\dot{x}^a)$ we have the following system of coupled ordinary
dif\/ferential equations for $\widehat{\gamma}$:
\begin{equation}\label{parallel transport}
x^a(t+\varepsilon)=x^a(t)+\varepsilon \dot{\gamma}^a(t),\qquad
\dot{x}^a(t+\varepsilon)=\dot{x}^a(t)-\varepsilon \dot{\gamma}^c(t)\,\dot{x}^b(t) \varGamma_{bc}^{a}\bigl(x(t)\bigr).
\end{equation}
Here $\dot{\gamma}^a(t)$ stands for the components of the tangent vector $\dot{\gamma}(t)\in T_{\gamma(t)}M$ to $\gamma$
with respect to the base coordinate basis $\{\partial_{x^a}|_{\gamma(t)}\}$.

Specifying the starting point for $\widehat{\gamma}$ in $TM$, i.e.\ the initial conditions for the above dif\/ferential system:
\[
\Bigl\{x^a(0)=x^a(\gamma(0)),\ \dot{x}^a(0)=v^a\Bigr\}\ \Longleftrightarrow\ v=v^a\partial_{x^a}\Bigr|_{\gamma(0)}
\]
we are able to perform a parallel transport of the vector $v\in T_{\gamma(0)}M$ along $\gamma$. Parallel transported
vector is the solution of (\ref{parallel transport}) at the f\/inal time $t=1$, i.e.\ $\dot{x}^a(1)\partial_{x^a}|_{\gamma(1)}$.

For a f\/ixed connection, there is a special class of curves that are called geodesics. Curve $\gamma:[0,1]\rightarrow M$
is a geodesic line on $M$, if parallel transport of $\dot{\gamma}(0)$ along $\gamma$ coincides at each time $t$
with its actual tangent vector $\dot{\gamma}(t)$.
In other words, being in a local chart on $M$, the geodesic $\gamma$ satisf\/ies the system of second
order dif\/ferential equations:
\[
\ddot{x}^a\bigl(\gamma(t)\bigr)=-\dot{x}^c\bigl(\gamma(t)\bigr) \dot{x}^b\bigl(\gamma(t)\bigr) \varGamma_{bc}^{a}\bigl(x\bigl(\gamma(t)\bigr)\bigr).
\]

With any af\/f\/ine connection one can associate its torsion and curvature (for more details see~\cite{geometers}).
Torsion of two vectors $u,v\in T_pM$ is a vector $\mathbf{T}_p (u,v)$
at point $p$, which closes an inf\/initesimal geodesic parallelogram framed on
$u$, $v$ up to second order.
Thus
\[
\mathbf{T}_p (u,v)=u^{b}v^{c} \mathbf{T}^a_{bc}(p) \partial_{x^a}\Bigr|_{p}=u^{b}v^{c}\Bigl\{\varGamma_{cb}^{a}-\varGamma_{bc}^{a}\Bigr\}(p) \partial_{x^a}\Bigr|_{p} .
\]
It is clear that torsion is an element of
$\Gamma\bigl(\bigwedge^2T^*M\otimes TM\bigr)$, i.e.\ it is skew-symmetric in the subscript indices.

Curvature associated with tangent vectors $u,v\in T_pM$ is a linear map, $(1,1)$-tensor,
$\mathbf{R}_p (u,v):T_pM\rightarrow T_pM$.
Again, analogically as before, the curvature measures how a vector $w$ changes, up to second order, after parallel
transport along an inf\/initesimal closed parallelogram spanned on vectors $u$ and $v$. Concisely,
\[
\mathbf{R}_p\,(u,v)(w)=w^{b}u^{c}v^{d}\,\mathbf{R}^a_{bcd}(p) \partial_{x^a}\Bigr|_{p}=w^{b}u^{c}v^{d}\Bigl\{\varGamma^{a}_{bd, c}-\varGamma^{a}_{bc, d}+\varGamma^{i}_{bd}\varGamma^{a}_{ic}-\varGamma^{i}_{bc}\varGamma^{a}_{id}\Bigr\}(p) \partial_{x^a}\Bigr|_{p}.
\]

We shall now describe how this is related to the Lagrange mechanics. The kinetic energy tensor $g$, due to
its symmetry, can be equivalently interpreted as a quadratic function on $TM$ w.r.t.\ f\/iber (doted) coordinates.
Using a local chart, one has:
\[
g=\tfrac{1}{2} g_{ab}(x)\dot{x}^a\dot{x}^b .
\]
Lets def\/ine the horizontal distribution at tangent bundle point $(p\,,v)$ by
\[
\mathrm{Hor}_{(p,v)}:=\Bigl\{w\bigr|_{(p,v)}=w^a\partial_{x^a}\bigr|_{(p,v)}^{\mathrm{Hor}}\in T_{(p,v)}(TM);\, w\bigr|_{(p,v)}(g)=0\Bigr\}\,.
\]
Requiring also that it is torsionless, one gets a system of equations for the Riemann--Levi-Civita connection:
\[
\partial_{x^c}(g_{ab})\equiv g_{ab, c}=g_{ai}\varGamma_{bc}^{i}+g_{bi}\varGamma_{ac}^{i},\qquad
0=\varGamma_{cb}^{a}-\varGamma_{bc}^{a} .
\]
Regularity of the $(m\times m)$ matrix $g$ ensures the unique solution of the above algebraic equations:
\[
\varGamma_{bc}^{a}=\tfrac{1}{2}\,g^{ai}\bigl\{g_{ib,\,c}+g_{ic,\,b}-g_{bc,\,i}\bigr\}\equiv\left\{{a \atop b\,c}\right\}_{_\mathrm{RLC}}
\]
and the existence of the canonical (musical\footnote{$\sharp_g$ symbolizes the index raising with
the help of dual (inverse) matrix $g^{ai}(\cdot g_{ib}=\delta^{a}_{b})$, for the index lowering with $g_{ai}$ there
is symbol $\flat_g$.})
isomorphism $\sharp_g:T^*M\rightarrow TM$. Therefore the external strength
described by the co-vectorial f\/ield $Q=Q_a(x,\dot{x}) dx^a$ can be turned into a force vector f\/ield
$\sharp_g(Q)=(g^{ab}Q_{b})\partial_a=F^a\partial_a=F$.

Suppose that a mechanical system at time $t$ is surrounded at given point $p\in M$ and possesses the velocity
$v\in T_{p}M$. Then the new position and velocity after the inf\/initesimal time interval~$\varepsilon$ is
determined by the Lagrange dynamical vector f\/ield evaluated at $(p,v)$
\begin{equation}\label{Lagrange}
\mathcal{L}\Bigr|_{(p,v)}=v\Bigr|_{(p,v)}^{\mathrm{Hor}}+ F\Bigr|_{(p,v)}^{\mathrm{Ver}}=
v^a\partial_{x^a}\Bigr|_{(p,v)}^{\mathrm{Hor}}+\ F^a(p\,,v)\partial_{\dot{x}^a}\Bigr|_{(p,v)}\in T_{(p,v)}(TM).
\end{equation}
Expressing this dynamics in local coordinates $(x^a,\dot{x}^a)$, one gets the following system of
dif\/fe\-ren\-tial equations:
\begin{gather*}
x^a(t+\varepsilon)=x^a(t)+\varepsilon\dot{x}^a(t),\\
\dot{x}^a(t+\varepsilon)=\dot{x}^a(t)-\varepsilon \dot{x}^c(t) \dot{x}^b(t) \varGamma_{bc}^{a}\bigl(x(t)\bigr)+\varepsilon  F^a\bigl(x(t),\dot{x}(t)\bigr) .
\end{gather*}
It is straightforward to see that these equations are exactly the Euler--Lagrange equations from the
beginning of this introductory paragraph. One just needs to replace the general $\varGamma_{bc}^{a}$ by
the RLC Christof\/fel symbols $\left\{{a \atop b\,c}\right\}_{_\mathrm{RLC}}$.

\section{Noncommutative Lagrange mechanics}

In the previous section we have seen that the mechanical system would evolve geodesically. The only disturbance of such
motion is caused by the presence of external forces. Thus we can conclude that the pivotal object def\/ining the
``theory'' consists in its internal kinetic energy tensor $g$ (lets call it a metric). States of the mechanical system
are labeled by points of the tangent bundle of the underlying conf\/iguration space $M$ (no internal
degrees of freedom are considered).

``Noncommutativity'' is an internal property of $M$ coming a priori from the Nature. In general it is
specif\/ied by nonconstant Poisson brackets def\/ined on $C^\infty(M)$, i.e.\ by a certain Poisson bi-vector f\/ield
$\Pi\in\Gamma(\bigwedge^2 TM)$.
Let us recall that locally:
\[
\Pi=\tfrac{1}{2} \Pi^{ab}(x)\bigl\{\partial_{x^a}\otimes\partial_{x^b}-\partial_{x^b}\otimes\partial_{x^a}\bigr\}
\
\Longleftrightarrow
\
\{\mathrm{f},\mathrm{h}\}:=\Pi(d\mathrm{f},d\mathrm{h})\equiv\mathrm{f}_{,a}\Pi^{ab}\mathrm{h}_{,b}.
\]
In terms of the bi-vector $\Pi$ the Jacobi identity is equivalent to the vanishing of the Schouten brackets of
the bi-vector $\Pi$ with itself, i.e.
\[
0=[\Pi,\Pi]_{_\mathrm{Schouten}}
\
\Longleftrightarrow
\
\Bigl\{\Pi^{ai} (\Pi^{bc})_{,i}+\mbox{cyclic permutation in }(a,b,c)\Bigr\}=0.
\]

The presence of the Poisson structure on $M$ enables us to quantize the algebra of functions on~$M$ (see~\cite{deformers}).
Quantization is a formal one parameter deformation of the ordinary pointwise product on $C^\infty(M)$. Resulting
noncommutative associative star-product algebra $\bigl(C^\infty(M)[[\hbar]],\star_{_\hbar}\bigr)$ satisf\/ies (in
the semiclassical regime):
\[
\lim\limits_{\hbar\rightarrow 0} \frac{1}{\hbar}\,\bigl(\mathrm{f} \star_{_\hbar}  \mathrm{h} - \mathrm{h} \star_{_\hbar} \mathrm{f}\bigr)=\Pi(d\mathrm{f},d\mathrm{h}) .
\]
What physicists are quite often doing when discussing the noncommutative theories (see e.g.~\cite{douglas-szabo}),
is a replacement of the ordinary pointwise product in the governing action
(Lagrangian density) by an appropriate star-product $\star_{_\hbar}$ and going to quantum theory.
The techniques and dictionary used for analyzing of the appearing NC-problems are classical (commutative)
ones.

If the noncommutativity is manifestly present in Nature, then it should somehow more conceptually af\/fect matters
also on the classical level. Not only in partial changes in the guiding Lagrangian and/or Hamiltonian, but directly,
modifying the classical equations of motion. These are rather more fundamental than the Lagrangian and/or Hamiltonian
themselves. Therefore what I am going to do below, is to pursue this direction.

Suppose that the classical ``theory'' is specif\/ied by the metric $g$ and occupies the noncommutative conf\/iguration
space $(M, \Pi)$. Can one use those ingredients to create something new? Let us consider the $(1,1)$-tensor
lowering one index of $\Pi$ with the help of $g$ (this I called the minimal coupling of the ``theory'' with
the ``noncommutativity''), i.e.
\[
\mathrm{Contr}(g\otimes\Pi)=dx^a g_{ai}(x) \Pi^{ib}(x)\partial_{x^b}=dx^a \Pi^{b}_{a}(x)\partial_{x^b}=\flat_g\Pi .
\]
With any $(1,1)$-tensor $A$ one can associate the special $(1,2)$-tensor called its Nijenhuis torsion~\cite{nijenhuis}.
It is def\/ined as follows:
\[
N_{A}(X,Y):=\bigl[A(X),A(Y)\bigr]-A\bigl(\bigl[A(X),Y\bigr]+\bigl[X,A(Y)\bigr]-A\bigl(\bigl[X,Y\bigr]\bigr)\bigr).
\]
Here the arguments $X$ and $Y$ are arbitrary vector f\/ields on $M$ and $[\,,\,]$ stands for their commutator. Checking
$C^\infty(M)$-linearity and exhibiting its properties w.r.t.\ argument interchange, one immediately concludes
that $N_{A}\in\Gamma(\bigwedge^2 T^*M\otimes TM)$, i.e.\ it is skew-symmetric in the subscript co-vectorial indices.

What do we get when substituting $\flat_g\Pi$ for $A$? Nijenhuis torsion of $\flat_g\Pi$ is expressed in the local
coordinate chart on $M$ as follows:
\begin{equation}\label{Nijenhuis}
N_{\flat_g\Pi}=dx^b\wedge
dx^c\,\tfrac{1}{2} \Bigl\{\Pi_{b}^{i} (\Pi^{a}_{c})_{,i}+\Pi^{a}_{i}(\Pi^{i}_{b})_{,c}\Bigr\}\partial_{x^a}.
\end{equation}

The $(1,2)$-tensor $N_{\flat_g\Pi}$ appears on $M$ very naturally. It ref\/lects the properties of both entering entities:
``theory'' $g$ and ``noncommutativity'' $\Pi$. Therefore it is quite reasonable to postulate: \emph{the affine
connection for the NC-classical mechanics is the metric $g$ connection, whose torsion is $N_{\flat_g\Pi}$}. This
asks us to solve the modif\/ied Riemann--Levi-Civita equations:
\[
g_{ab,\,c}=g_{ai}\varGamma_{bc}^{i}+g_{bi}\varGamma_{ac}^{i},\qquad
\mathbf{T}_{bc}^a=\varGamma_{cb}^{a}-\varGamma_{bc}^{a},
\]
with
\begin{equation}\label{NC-torsion}
\mathbf{T}_{bc}^a=\Pi_{b}^{i} (\Pi_{c}^{a})_{,i}-\Pi_{c}^{i} (\Pi_{b}^{a})_{,i}+\Pi^{a}_{i}(\Pi^{i}_{b})_{,c}-\Pi^{a}_{i}(\Pi^{i}_{c})_{,b}=\bigl(N_{\flat_g\Pi}\bigr)_{bc}^a.
\end{equation}
The f\/inal formulae for $\varGamma$'s are given as:
\[
\varGamma_{bc}^{a}=\tfrac{1}{2}g^{ai}\bigl\{g_{ib,c}+g_{ic,b}-g_{bc,i}\bigr\}
-\tfrac{1}{2}g^{ai}\bigl\{g_{ij}\mathbf{T}^{j}_{bc}+g_{bj}\mathbf{T}^{j}_{ci}+g_{cj}\mathbf{T}^{j}_{bi}\bigr\}=
\left\{{a \atop b\,c}\right\}_{_\mathrm{RLC}}-\mathbf{K}^{a}_{bc}.
\]
The dynamics of a given state $(p=\mbox{position},v=\mbox{velocity})$ at time $t$ is governed by
the inf\/initesimal f\/low of the Lagrange dynamical vector f\/ield (\ref{Lagrange}) evaluated at $(p\,,v)\in TM$.
Let us stress that its horizontal part is now specif\/ied by the modif\/ied set of Christof\/fel symbols. Apart from the
standard (commutative) RLC-term there appears the additional tensorial part~$\mathbf{K}^{a}_{bc}$ called the
\emph{contortion}. Thus the NC-Lagrange equations can be written as follows:
\begin{equation}\label{NC-Lagrange}
\ddot{x}^a+\dot{x}^c\dot{x}^b\left\{{a \atop b\,c}\right\}_{_\mathrm{RLC}} = F^a+\dot{x}^c\dot{x}^b \mathbf{K}^{a}_{bc},
\qquad a=1,\dots, m.
\end{equation}

Using the optics of the standard Lagrange mechanics: the ``noncommutativity'' can be treated as an
additional ``internal'' force. Its origin comes from the presence of the nonzero contortion. To reveal
its physical consequences let us recall what geometrically the contortion is responsible for. Parallel transport
(\ref{parallel transport}) along $\gamma$ determined by
$\varGamma^{a}_{bc}=\left\{{a \atop b\,c}\right\}_{_\mathrm{RLC}} -\mathbf{K}^{a}_{bc}$ can be split into two steps:
\[
v^a\longmapsto v_{\mathrm{r}}^a=v^a+\varepsilon v^b\bigl[\mathbf{K}^{a}_{bc}\dot{\gamma}^c\bigr]+o(\varepsilon)
\longmapsto v^a_{\mathrm{r}}-\varepsilon \dot{\gamma}^c v_{\mathrm{r}}^b\left\{{a \atop b\,c}\right\}_{_\mathrm{RLC}}+o(\varepsilon).
\]
The f\/irst arrow is an inf\/initesimal linear transformation determined by $\mathbf{K}(\cdot,\dot{\gamma})$ that takes
place in the same tangent space as the initial vector $v$. The second one is just the standard Levi-Civita parallel transport
of $v_{\mathrm{r}}$. Since the considered connection is compatible with the metric, both $v$ and $v_{\mathrm{r}}$ have
the same lengths and therefore $\mathbf{K}(\cdot,\dot{\gamma})$
is a generator of some rotation (it depends on the tangent vector $\dot{\gamma}$ to the transporting curve $\gamma$).
Thus the ef\/fect of additional noncommutative ``Lorentz-like'' force is an inf\/initesimal rotation of actual velocity
that precedes the next ``commutative'' RLC-step. Let us recall that mechanics with non zero torsion is the
subject of study at \cite{shabanov}, similar problem from the general relativity point of view is analyzed in~\cite{hehl}.

\medskip

\noindent\textbf{Example.} Suppose that $M$ is ordinary two-dimensional plane, which is endowed by the f\/lat metric
$g=\frac{1}{2}\,\{dx\otimes dx+dy\otimes dy\}$. Noncommutativity on $M$ is imposed by a general Poisson bracket:
\[
\{x,y\}=\Theta(x,y),\qquad \mbox{here} \quad \Theta(x,y) \ \mbox{is any (well-behaved) function on} \ \mathbb{R}^2.
\]
The dynamics of a free particle is then governed by the NC-force, whose Cartesian components are given as follows:
\[
F_{_\mathrm{NC}}^x=\dot{y} \Theta \{\dot{x}\Theta_{,y}-\dot{y}\Theta_{,x}\},\qquad \mbox{and}\qquad
F_{_\mathrm{NC}}^y=-\dot{x}\Theta\{\dot{x}\Theta_{,y}-\dot{y}\Theta_{,x}\}.
\]

\section{Concluding remarks}

\noindent $\bullet$ The dynamical system we have considered was an autonomous
one. But in a general metric, noncommutativity, as well as the external strengths, could be explicitly time
dependent. In that case, one should work with the connection on the f\/irst jet-extension of $M\times\mathbb{R}$.
Emerging complications are mainly technical.

\vskip 2mm

\noindent $\bullet$ The Lorentz-like force specifying the coupling of the noncommutative background
with the system under consideration comes from pure geometrical considerations. Very similar magnetic-like background
ef\/fect of the noncommutativity emerges also in the standard Moyal plane approach, see \cite{chujkovia1,chujkovia2,chujkovia3}.
Here it is due to the af\/f\/ine connection with the nonzero contortion. There it is so, because
\[
[x_a,x_b]=i\Theta_{ab}\qquad (a,b=1,2)
\]
in two-dimensions provides an ef\/fective coupling with the axial magnetic f\/ield
$\mathrm{B}_3=\Theta_{12}$. Both approaches to the noncommutative mechanics are
therefore ideologically independent and physically inequivalent. This can be easily seen for example, if
$\Theta_{ab}=\mathrm{const}\,\epsilon_{ab}$. Then there is a~nonzero magnetic f\/ield $\mathbf{B}=(0,0,\mathrm{B}_3)$, but
since $\mathbf{K}^{c}_{ab}$ depends on derivatives of $\Theta_{ab}$, the contortion term is zero.

\vskip 2mm

\noindent $\bullet$ The question of symmetries of equations of motion has an obvious solution. The base vector
f\/ield $V=V^i(x)\partial_{x^i}\in \Gamma(TM)$ is a generator of symmetry of the considered dynamical system specif\/ied
by (\ref{NC-Lagrange}), if its complete (natural, also called horizontal) tangent bundle lift
\[
V^{\mathrm{c}}=V^a(x)\partial_{x^a}+\dot{x}^b \bigl(V^a(x)\bigr)_{,b} \partial_{\dot{x}^a},
\]
commutes with the Lagrange dynamical f\/ield $\mathcal{L}$. Let me remind the reader that $\mathcal{L}$ and $V^{\mathrm{c}}$
are both the vector f\/ields def\/ined on the tangent bundle~$TM$.

\vskip 2mm

\noindent $\bullet$ The additional contortion dependent force emerging in (\ref{NC-Lagrange}) is
not potential-generated neither generalized potential-generated, i.e.\ there does not exist a tangent bundle potential
function~$U(x,\dot{x})$, such that
\[
\mathbf{K}_{abc} \dot{x}^b\dot{x}^c\equiv g_{ai} \mathbf{K}^{i}_{bc} \dot{x}^b\dot{x}^c=-\frac{\partial U}{\partial x^a}+\frac{d}{dt}\left(\frac{\partial U}{\partial \dot{x}^a}\right).
\]
The proposed NC-dynamics is therefore not derivable from some ``non\-com\-mu\-ta\-ti\-ve'' Lagrangian and/or Hamiltonian
function in the obvious fashion. The question about the possible quantization thus remains open (at least for the author
of this paper). Since we do not have Hamiltonian~$H$, we can not write:
\[
\dot{x}=\{x,H\}_{\star_{_\hbar}},\qquad \dot{p}=\{p,H\}_{\star_{_\hbar}},\qquad \mbox{where}\qquad
\{\mathrm{f},\mathrm{h}\}_{\star_{_\hbar}}:=\mathrm{f}\star_{_\hbar} \mathrm{h} - \mathrm{h}\star_{_\hbar} \mathrm{f}
\]
and the Moyal--Kontsevich $\star_{_\hbar}$-product is def\/ined with respect to the given Poisson structure $\Pi$
which specif\/ies the noncommutative background under consideration.

\vskip 2mm

\noindent $\bullet$ Suppose that the external forces are potential-generated, then the classical energy can
be introduced as standardly:
\[
\mathrm{E}=\tfrac{1}{2} g_{ab}(x)\dot{x}^a\dot{x}^b+U(x)=T+U.
\]
Verif\/ication that $\mathrm{E}$ is conserved by the NC-dynamics is straightforward\footnote{Concisely, take
the scalar product of (\ref{NC-Lagrange}) with the actual velocity $\dot{x}^a$ and use the mentioned
``Lorentz-likeness'' of the additional contortion term $\dot{x}^a\mathbf{K}_{abc}\dot{x}^b\dot{x}^c=0=
g(v,\mathbf{K}(v,v))$, then
\[
g_{ab} \dot{x}^a\Bigr[\ddot{x}^b+\dot{x}^i\dot{x}^j\Bigl\{{b \atop j\,i}\Bigr\}_{_{\mathrm{RLC}}} \Bigr]=
g_{ab} \dot{x}^a\ddot{x}^b+\tfrac{1}{2} g_{ab,i}\dot{x}^a\dot{x}^b\dot{x}^i\bigl(=\dot{T}\bigr)
\overset{(\ref{NC-Lagrange})}{=}-\dot{x}^a U_{,a}+0=-\dot{U}\ \Longrightarrow\ \dot{\mathrm{E}}=0.
\]}.

\vskip 2mm

\noindent $\bullet$ The dynamics in the Hamiltonian ``picture'' is just a simple combination of cotangent bundle parallel
transport\footnote{Concisely, parallel transport of a co-vector $\alpha\in T^*_pM$ along $\gamma$
is ef\/fectively def\/ined via its $g$-dual, i.e.\ the vector $\sharp_g\alpha\in T_pM$. Transporting it and taking its inverse
dual at a second terminal point of $\gamma$ one gets parallel transport of the co-vector $\alpha$.} and vertical
lift of the co-vectorial strength f\/ield $Q$. Suppose that at some time~$t$ the mechanical system occupies a cotangent
bundle state
$(p=\mbox{position},\alpha=\mbox{mo\-men\-tum})$, then at inf\/initesimally close time the new $T^*M$-position is
determined by the f\/low of the Hamilton dynamical f\/ield:
\[
\mathcal{H}\Bigr|_{(p,\alpha)}=\bigl(\sharp_g\alpha\bigr)\Bigr|_{(p,\alpha)}^{\mathrm{Hor}}+ Q\Bigr|_{(p,\alpha)}^{\mathrm{Ver}}
\in T_{(p,\alpha)}(T^*M).
\]

\vskip 2mm

\noindent $\bullet$  The conf\/iguration (Riemannian) space $(M,g)$ could be at any time replaced by
a pseudo-Riemannian
space-time manifold. Then one can easily impose ``non\-com\-mu\-ta\-ti\-vi\-ty'' within gene\-ral relativity modifying
the underlaying metric connection by non zero torsion terms. General relativity with torsion is analyzed at
\cite{hehl}.

\subsection*{Acknowledgements}

This research was supported in part by M\v SMT \v CR grant LC06002, ESF projects: JPD 3 2005/NP1-013 and JPD 3BA 2005/1-034 and VEGA Grant
1/3042/06. The author is thankful to Pavel Exner, Miloslav Znojil and Jaroslav Dittrich for their kindness and
hospitality during the author's short-time fellowship at the Doppler Institute in the autumn 2006.

\pdfbookmark[1]{References}{ref}
\LastPageEnding

\end{document}